%% file: ms.tex
\documentclass[sigconf,screen]{acmart}

\usepackage[utf8]{inputenc}
\usepackage[T1]{fontenc}
\usepackage{microtype}

\usepackage{color,xcolor}
\usepackage{xspace}
\usepackage{multirow,tabularx,adjustbox}
\usepackage[para]{threeparttable}
\usepackage{subcaption}
\usepackage{tikz}
\usepackage{enumitem}
\usepackage{balance}
\usetikzlibrary{shapes,arrows}
\graphicspath{{./figures/}}

\definecolor{riversideblue}{HTML}{2d6cc0}

\newcommand{\mrCell}[3]{\multirow{#1}{*}{\parbox{#2cm}{\centering #3}}}
\newcommand{\mcCell}[3]{\multicolumn{#1}{c}{\parbox{#2cm}{\centering #3}}}

\newif\ifdraft\drafttrue

  {\begin{list}{$\bullet$}%
     {\setlength{\parsep}{0pt}%
      \setlength{\topsep}{0pt}%
      \setlength{\leftmargin}{0.75pc}%
      \setlength{\itemsep}{2pt}}}%
  {\end{list}}

\usepackage{listings}
\definecolor{KWColor}{rgb}{0.37,0.08,0.25}
\definecolor{CommentColor}{rgb}{0.12,0.38,0.18}
\definecolor{StringColor}{rgb}{0.06,0.10,0.98}
\definecolor{darkred}{rgb}{0.75,0,0}
\definecolor{lightgrey}{rgb}{0.8,0.8,0.8}

\lstdefinestyle{Eclipse}{
  xleftmargin=0pt,
  basicstyle=\ttfamily\small,
  commentstyle=\color{CommentColor}\ttfamily\small,
  stringstyle=\color{StringColor},
  keywordstyle=\color{KWColor}\bfseries,
  escapeinside={/*@}{@*/}
}

\newcommand{\code}[1]{\lstinline{#1}}

\def\Snospace~{\S{}}

\tikzstyle{decision} = [diamond, draw 
    text width=4.5em, text badly centered, node distance=3cm, inner sep=0pt]
\tikzstyle{block} = [rectangle, draw, 
    text width=27em, text centered, rounded corners, minimum height=2em]
\tikzstyle{line} = [draw, -latex']
\tikzstyle{cloud} = [draw, ellipse, node distance=2cm,
    text width=5em, minimum height=2em]

\newcommand{\nullaway}{\textsc{NullAway}\xspace}
\newcommand{\uberfull}{Uber Technologies Inc.\xspace}
\newcommand{\uber}{Uber\xspace}
\newcommand{\cfnullness}{\textsc{CFNullness}\xspace}
\newcommand{\initannot}{\code{@Initializer}\xspace}
\newcommand{\nonnull}{\code{@NonNull}\xspace}
\newcommand{\nullable}{\code{@Nullable}\xspace}

\newcommand{\mypara}[1]{\noindent{\textbf{#1}}}

\setcopyright{acmlicensed}
\acmPrice{15.00}
\acmDOI{10.1145/3338906.3338919}
\acmYear{2019}
\copyrightyear{2019}
\acmISBN{978-1-4503-5572-8/19/08}
\acmConference[ESEC/FSE '19]{Proceedings of the 27th ACM Joint European Software Engineering Conference and Symposium on the Foundations of Software Engineering}{August 26--30, 2019}{Tallinn, Estonia}
\acmBooktitle{Proceedings of the 27th ACM Joint European Software Engineering Conference and Symposium on the Foundations of Software Engineering (ESEC/FSE '19), August 26--30, 2019, Tallinn, Estonia}


\begin{document}

\title{\nullaway : Practical Type-Based Null Safety for Java}

\author{Subarno Banerjee}
\affiliation{%
  \institution{University of Michigan}
  \city{Ann Arbor}
  \state{MI}
  \country{USA}}
\email{subarno@umich.edu}

\author{Lazaro Clapp}
\affiliation{%
  \institution{Uber Technologies, Inc.}
  \city{San Francisco}
  \state{CA}
  \country{USA}}
\email{lazaro@uber.com}

\author{Manu Sridharan}
\affiliation{%
  \institution{University of California, Riverside}
  \city{Riverside}
  \state{CA}
  \country{USA}}
\email{manu@cs.ucr.edu}

\input{sections/abstract}


\maketitle

\input{sections/intro}
\input{sections/design}

\input{sections/implementation}
\input{sections/evaluation}
\input{sections/related}
\input{sections/conclusion}

\balance
\bibliographystyle{ACM-Reference-Format}
\bibliography{refs}

\end{document}

%% file: sections/abstract.tex

\begin{abstract}
\code{NullPointerException}s (NPEs) are a key source of crashes in modern Java programs.  Previous work has shown how such errors can be prevented at compile time via code annotations and pluggable type checking.  However, such systems have been difficult to deploy on large-scale software projects, due to significant build-time overhead and / or a high annotation burden.  This paper presents \nullaway, a new type-based null safety checker for Java that overcomes these issues.  \nullaway has been carefully engineered for low overhead, so it can run as part of every build.  Further, \nullaway reduces annotation burden through targeted unsound assumptions, aiming for \emph{no false negatives in practice} on checked code.  Our evaluation shows that \nullaway has significantly lower build-time overhead ($1.15\times$) than comparable tools ($2.8$-$5.1\times$).  Further, on a corpus of production crash data for widely-used Android apps built with \nullaway, remaining NPEs were due to unchecked third-party libraries ($64\%$), deliberate error suppressions ($17\%$), or reflection and other forms of post-checking code modification ($17\%$), \emph{never} due to \nullaway's unsound assumptions for checked code.
\end{abstract}

%% file: sections/intro.tex

\section{Introduction}


\code{NullPointerException}s (NPEs), caused by a dereference of \code{null}, are a frequent cause of crashes in modern Java applications.  Such crashes are nearly always troublesome, but they are particularly problematic in mobile applications.  Unlike server-side code, where a bug fix can be deployed to all users quickly, getting a fixed mobile app to users' devices can take days to weeks, depending on the app store release process and how often users install updates.  Due to the severity of null-dereference errors, recent mainstream languages like Swift~\cite{swift-lang} and Kotlin~\cite{kotlin-lang} enforce null safety as part of type checking during compilation.


Previous work has added type-based null safety to Java, via code annotations and additional type checking~\cite{PapiACPE2008,DietlDEMS2011,tool-eradicate}.  With this approach, developers use \nullable and \nonnull code annotations to indicate whether entities like fields, parameters, and return values may or may not be \code{null}.  Given these annotations, a tool checks that the code is null safe, by ensuring, e.g., that \nullable expressions are never de-referenced and that \code{null} is never assigned to a \nonnull variable.  Previous work has shown this approach to be an effective way to prevent NPEs~\cite{PapiACPE2008,DietlDEMS2011}.


Despite their effectiveness, previous type-based null safety tools for Java suffered from two key drawbacks.  First, the build-time overhead of such tools is quite high.  Our experimental evaluation showed the two best-known tools to have average overheads of $2.8\times$ and $5.1\times$ respectively (see \autoref{sec:eval}) compared to regular compilation.  For a seamless development experience, a null safety tool should run every time the code is compiled, but previous tool overheads are too high to achieve this workflow without excessive impact on developer productivity.  Second, some previous tools prioritize soundness, i.e., providing a strong guarantee that any type-safe program will be free of NPEs.  While this guarantee is appealing in principle, it can lead to significant additional annotation burden for developers, limiting tool adoption.


To address these drawbacks, we have developed \nullaway, a new tool for type-based null safety for Java.  \nullaway runs as a plugin to the Error Prone framework~\cite{DBLP:conf/scam/AftandilianSPK12}, which provides a simple API for extending the Java compiler with additional checks.  The core of \nullaway includes features of previous type-based null safety tools, including defaults that reduce the annotation burden and flow-sensitive type inference and refinement~\cite{PapiACPE2008}.  \nullaway includes additional features to reduce false positives, such as support for basic pre-/post-conditions and for stream-based APIs (\autoref{sec:other-features}).  

\nullaway is carefully engineered and regularly profiled to ensure low build-time overhead.  We built \nullaway at \uberfull (\uber), and have run it as part of all our Android builds (both on continuous integration servers and developer laptops) for over two years.  At \uber, \nullaway replaced another tool which, due to performance limitations, ran only at code review time.  Running \nullaway on all builds enabled much faster feedback to developers.



Regarding soundness, \nullaway aims to have \emph{no false negatives in practice for code that it checks}, while reducing the annotation burden wherever possible.  \nullaway's checks to ensure \nonnull fields are properly initialized (\autoref{sec:init-checking}) are unsound, but also require far fewer annotations than a previous sound checker~\cite[\S 3.8]{cf-manual}.  Similarly, \nullaway unsoundly assumes that methods are pure, i.e., side-effect-free and deterministic (\autoref{sec:side-effects}).  In both cases, we have validated that neither source of unsoundness seems to lead to real-world NPEs for \uber's Android apps, based on crash data from the field.

For usability, \nullaway uses an \emph{optimistic} handling of calls into unchecked code, though such handling can lead to uncaught issues.  Modern Java projects often depend on numerous third-party libraries, many of which do not yet contain nullability annotations on their public APIs.  Maximum safety requires a \emph{pessimistic} modeling of such libraries, with worst-case assumptions about their nullness-related behavior.  Pessimistic assumptions lead to a large number of false positive warnings, in our experience making the tool unusable on a large code base.  Instead, \nullaway treats calls into unchecked code \emph{optimistically}: it assumes that such methods can always handle \code{null} parameters and will never return \code{null}.\footnote{Optimistic handling is also used for overriding methods from unchecked packages; see \autoref{sec:unannotated}.}  Additionally, \nullaway includes mechanisms for custom library modeling and leveraging nullability annotations when they are present.  Overall, \nullaway's handling of unchecked code is practical for large code bases while providing mechanisms for additional safety where needed.

We performed an extensive experimental evaluation of \nullaway, on $18$ open-source projects totaling ${\sim}164K$ lines of code and on ${\sim}3.3M$  lines of production code from widely-used Android apps developed at \uber.  We observed that \nullaway introduced an average of $15\%$ overhead to build times on the former ($22\%$ on the later), significantly lower than previous tools.  Further, a study of one month of crash data from \uber showed that NPEs were uncommon, and that nearly all remaining NPEs were due to interactions with unchecked code, suppression of \nullaway warnings, or post-checking code modification.  \emph{None} of the NPEs were due to \nullaway's unsound assumptions for checked code.  Finally, the evaluation confirmed that removing these unsound assumptions leads to significantly more warnings for developers.

\nullaway is freely available and open source.  It has more than 2,500 stars on GitHub~\cite{tool-nullaway}, and has been adopted by a number of other companies and open-source projects, further validating its usefulness.  We believe that \nullaway's design and tradeoffs provide a useful template for future type systems aiming to prevent crashes in large-scale code bases.

\mypara{Contributions} This paper makes the following contributions:
\begin{itemize}
\item We describe the design of \nullaway's type system, tuned over many months to achieve no false negatives in practice for checked code with a reasonable annotation burden.  \nullaway includes a novel, carefully-designed initialization checking algorithm (\autoref{sec:init-checking}), an optimistic treatment of method purity (\autoref{sec:side-effects}), and a highly-configurable system for determining how to treat unchecked code (\autoref{sec:unannotated}).  Our evaluation showed that a checker without these unsound assumptions emitted many false positive warnings (\autoref{sec:eval-cfnullness-warnings}).
\item We present experiments showing that \nullaway's build-time overhead is dramatically lower than alternative systems, enabling NPE checking on every build (\autoref{sec:eval-performance}).
\item We analyze production crash data for a large code base built with \nullaway and show that on this data set, \nullaway achieved its goal of no false negatives for checked code, as remaining NPEs were primarily caused by third-party libraries and warning suppressions (\autoref{sec:eval-uber}).
\end{itemize}



%% file: sections/design.tex

\section{Overview}\label{sec:overview}

In this section we give a brief overview of type-based nullability checking as implemented in \nullaway.  The core ideas of preventing NPEs via pluggable types are well known; see elsewhere~\cite{PapiACPE2008,DietlDEMS2011,cf-manual} for further background.

With type-based null checking, a type's nullability is expressed via additional qualifiers, written as annotations in Java. The \nonnull qualifier describes a type that excludes  \code{null}, whereas \nullable indicates the type includes \code{null}.  Given these additional type qualifiers, type checking ensures the following two key properties:
\begin{enumerate}
\item No expression of \nullable type is ever assigned to a location of \nonnull type.
\item No expression of \nullable type is ever dereferenced.
\end{enumerate}
Together, these properties ensure a program is free of NPEs, assuming objects have been properly initialized.  (We defer discussion of initialization checking to \autoref{sec:init-checking}.)

Consider the following simple example:
\begin{lstlisting}
void log(@NonNull Object x) {
    System.out.println(x.toString());
}
void foo() { log(null); }
\end{lstlisting}
Here, the parameter of \code{log} is \nonnull, so the call \code{log(null);} will yield a type error, as it violates property 1.\footnote{For type checking, ``assignments'' include parameter passing and returns at method calls.}  The developer could address this issue by changing the annotations on \code{log}'s parameter \code{x} to be \nullable.  But, \code{x} is dereferenced at the call \code{x.toString()}, which would yield another type error due to violating property 2.

One way the developer can make the code type check is to change the body of the \code{log} method as follows:
\begin{lstlisting}
void log(@Nullable Object x) {
  if (x != null) { System.out.println(x.toString()); } /*@ \label{li:flow-sensitive} @*/
}
\end{lstlisting}
The type checker proves this code safe via \emph{flow-sensitive type refinement} (to be discussed further at the end of this section); the checker interprets the null check and refines \code{x}'s nullness type to be \nonnull within the if-body, making the \code{toString()} call legal.

Types qualified with nullability annotations form a subtyping relationship, where \code{@NonNull C} is a subtype of \code{@Nullable C} for any class \code{C} \cite[Fig. 3]{PapiACPE2008}.  Hence, property 1 simply ensures assignment compatibility according to subtyping.

\mypara{Override Checking} \nullaway also ensures that method overrides respect subtyping, enforcing the standard function subtyping rules of covariant return types and contravariant parameter types~\cite{Pierce:2002:TPL:509043}.  Consider the following example:
\begin{lstlisting}
class Super {
  @NonNull Object getObj() { return new Object(); }
}
class Sub extends Super {
  @Nullable Object getObj() { return null; }
}
class Main {
  void caller() {
    Super x = new Sub();
    x.getObj().toString(); // NullPointerException! /*@ \label{li:fizzbuzz} @*/
  }
}
\end{lstlisting} 

\begin{sloppypar}
Since \code{x} has declared type \code{Super}, the declared target of \code{x.getObj()} on line~\ref{li:fizzbuzz} is \code{Super.getObj}.  This method has a \nonnull return type, making the \code{toString()} call legal.  However, this example crashes with an NPE, since overriding method \code{Sub.getObj} has \nullable return type.  To close this loophole, the checker must ensure covariance in return types, so a method with \nonnull return type cannot be overridden by one with \nullable return type.  Similarly, it must check for contravariant parameter types, so a method's \nullable parameter cannot be made \nonnull in an overriding method.
\end{sloppypar}

\mypara{Defaults} Annotating every field, parameter, and return value in a large code base would require a huge effort.  \nullaway uses the non-null-except-locals (NNEL) default from the Checker Framework~\cite{PapiACPE2008} to reduce the annotation burden.  Any unannotated parameter, field, or return value is treated as \nonnull, whereas the types of local variables are inferred (see below).  Beyond reducing annotation effort, this default makes code more readable (by reducing annotation clutter) and nudges the developer away from using \code{null} values, making the code safer.

\mypara{Flow-Sensitive Type Inference / Refinement} As in previous work, \nullaway automatically infers types for local variables in a flow-sensitive manner.  Beyond inspecting assignments, null checks in conditionals are interpreted to compute refined (path-sensitive) types where the condition holds.  E.g., at \autoref{li:flow-sensitive} of the previous \code{log} example, the type of \code{x} is refined to \nonnull inside the if-body, based on the null check.  \nullaway uses an access-path-based abstract domain~\cite{DBLP:conf/pldi/Deutsch94} to also track nullability of sequences of field accesses and method calls.  \autoref{sec:side-effects} describes how \nullaway's assumptions around method purity interact with type inference.

\mypara{Other Tools} Throughout this paper we discuss two other type-based null checking tools for Java: the Nullness Checker from the Checker Framework~\cite{PapiACPE2008,DietlDEMS2011}, which we refer to as \cfnullness for brevity, and Eradicate, available with Facebook Infer~\cite{tool-eradicate}.  Subsequent sections will detail how the checks performed by \nullaway and its overheads compare with \cfnullness and Eradicate.

\section{Initialization Checking}\label{sec:init-checking}

Beyond the checks shown in \autoref{sec:overview}, to fully prevent NPEs a nullness type checker must ensure that objects are properly initialized.  Sound type systems for checking object initialization have been a subject of much previous research~\cite{DBLP:conf/oopsla/FahndrichL03, DBLP:conf/oopsla/SummersM11, DBLP:conf/popl/QiM09, DBLP:conf/oopsla/FahndrichX07}.  In this section, we present \nullaway's approach to initialization checking.  Though unsound, our technique has a low annotation burden and has caught nearly all initialization errors in our experience at \uber.

\begin{figure}
\begin{lstlisting}
class InitExample {
  @NonNull Object f, g, h, k; /*@ \label{li:init-field-decls} @*/
  InitExample() { /*@ \label{li:init-constructor-start} @*/
    this.f = new Object(); /*@ \label{li:init-init-f} @*/
    this.g.toString(); // use before init /*@ \label{li:init-bad-g-read} @*/
    helper(); /*@ \label{li:init-helper-invoke} @*/
  }  /*@ \label{li:init-constructor-end} @*/
  private void helper() { /*@ \label{li:init-helper-start} @*/
    this.g = new Object(); /*@ \label{li:init-init-g} @*/
    this.h.toString(); // use before init /*@ \label{li:init-bad-h-read} @*/
  } /*@ \label{li:init-helper-end} @*/
  @Initializer public void init() { /*@ \label{li:init-init-method-start} @*/
    this.h = this.f; /*@ \label{li:init-init-h} @*/
    if (cond()) { this.k = new Object(); } /*@ \label{li:init-cond-init-k} @*/
  } /*@ \label{li:init-init-method-end} @*/
}
\end{lstlisting}
\caption{An example (with errors) to illustrate initialization checking.}
\label{fig:init-code-example}
\end{figure}

\autoref{fig:init-code-example} gives a code example we will use to illustrate our initialization checking.  We first describe how \nullaway checks that \nonnull fields are initialized (\autoref{sec:field-init}), then discuss checking for uses before initialization (\autoref{sec:use-before-init}), and then compare with \cfnullness and Eradicate (\autoref{sec:init-discussion}).

\subsection{Field Initialization}\label{sec:field-init}

\mypara{Initialization phase} Any \nonnull instance field must be assigned a non-null value by the end of the object's initialization phase.\footnote{For space, we elide discussion of \nullaway's handling of static field initialization; the techniques are roughly analogous to those for instance fields.}  We consider an object's initialization phase to encompass execution of a constructor, possibly followed by \emph{initializer methods}.  Initializer methods (or, simply, initializers) are methods invoked at the beginning of an object's lifecycle but after its constructor, e.g., overrides of \code{onCreate()} in Android \code{Activity} subclasses~\cite{activity-lifecycle}.  Field initialization may occur directly in constructors and initializers, or in invoked helper methods.

In \autoref{fig:init-code-example}, the \code{InitExample} class has four \nonnull fields, declared on \autoref{li:init-field-decls}.  \nullaway treats the \code{init()} method (lines~\ref{li:init-init-method-start}--\ref{li:init-init-method-end}) as an initializer, due to the \initannot annotation.  For a method annotated \initannot, \nullaway \emph{assumes} (without checking) that client code will \emph{always} invoke the method before other (non-initializer) methods in the class.  Note that the \code{InitExample} constructor invokes \code{helper()} at \autoref{li:init-helper-invoke} to perform some initialization.

\mypara{Checks} Given a class $C$ with constructors, initializers, and initializer blocks~\cite[\S 8.6]{jls9}, for each \nonnull field $f$ of $C$, \nullaway treats $f$ as properly initialized if any one of four conditions holds:
%
\begin{enumerate}
\item $f$ is initialized directly at its declaration; or
\item $f$ is initialized in an initializer block; or
\item $C$ has at least one constructor, and \emph{all} constructors initialize $f$; or
\item \emph{some} initializer in $C$ initializes $f$.
\end{enumerate}

For a method, constructor, or initializer block $m$ to initialize a field $f$, $f$ must \emph{always} be assigned a non-null value by the end of $m$.  This property can be determined using the same analysis used for flow-sensitive type inference (see \autoref{sec:overview}), by checking if the inferred type of \code{this.f} is \nonnull at the end of $m$.
\nullaway also allows for initialization to occur in a method that is \emph{always invoked} by $m$.  \nullaway determines if $m$ always invokes a method $n$ with two simple checks: (1) the call to $n$ must be a top-level statement in $m$ (not nested within a conditional or other block),\footnote{\nullaway currently (unsoundly) treats $n$ as always-invoked even if $m$ may return before invoking $n$.} and (2) $n$ must be \code{private} or \code{final}, to prevent overriding in subclasses.\footnote{\nullaway does not attempt to identify methods that are always invoked from a constructor or initializer through a chain of invocations more than one level deep; this has not led to false positive warnings in practice.}

For \autoref{fig:init-code-example}, \nullaway reasons about initialization as follows:
\begin{itemize}
\item \code{f} is properly initialized due to the assignment at \autoref{li:init-init-f}.
\item \code{g} is properly initialized, since the constructor always invokes \code{helper()} (\autoref{li:init-helper-invoke}), which assigns \code{g} (\autoref{li:init-init-g}).
\item \code{h} is properly initialized, since \initannot method \code{init()} assigns \code{h} (\autoref{li:init-init-h}).
\item Line~\ref{li:init-cond-init-k} only initializes \code{k} conditionally.  So, \nullaway reports an error that \code{k} is not properly initialized.
\end{itemize}

\subsection{Use before Initialization}\label{sec:use-before-init}

Within the initialization phase, a further check is required to ensure that \nonnull fields are not used before they are initialized.  Two such bad uses exist in \autoref{fig:init-code-example}: the read of \code{this.g} at \autoref{li:init-bad-g-read} and \code{this.h} at \autoref{li:init-bad-h-read}.  \nullaway performs a partial check for these bad uses.  Within constructors and initializers, \nullaway checks at any field use that the field is definitely initialized before the use.  This check again leverages the same analysis used for flow-sensitive type inference.  \nullaway must also account for fields that have been initialized before the analyzed method.  For example, the read of \code{this.f} at \autoref{li:init-init-h} of \autoref{fig:init-code-example} is safe, since \code{f} is initialized in the constructor, which runs earlier.  Similarly, \nullaway accounts for fields initialized in always-invoked methods before a read.

\nullaway's check is partial since it does not check field reads in methods invoked by constructors or initializers or guard against other leaking of the \code{this} reference during initialization.  So, while \nullaway reports a use-before-init error at \autoref{li:init-bad-g-read} of \autoref{fig:init-code-example}, it does not report an error for the uninitialized read at \autoref{li:init-bad-h-read}.  While handling certain cases like reads in always-invoked methods would be straightforward, detecting all possible uninitialized reads would be non-trivial and add significant complexity to \nullaway.  Uninitialized reads beyond those detected by \nullaway seem to be rare, so we have not yet added further checking.

\subsection{Discussion}\label{sec:init-discussion}


In contrast to \nullaway, \cfnullness aims for sound initialization checking.  The \cfnullness initialization checking system~\cite[\S 3.8]{cf-manual} (an extension of the Summers and M{\"{u}}ller type system~\cite{DBLP:conf/oopsla/SummersM11}) prohibits invoking any method on a partially-initialized object without additional developer annotations.  E.g., \cfnullness prohibits the call at \autoref{li:init-helper-invoke}, since \code{helper()} is not annotated as being able to operate during initialization.  It also lacks support for a direct analogue of the \initannot annotation.  As we shall show in \autoref{sec:eval} this strict checking leads to a number of additional false warnings.  \nullaway's checking is unsound, but it seems to catch most initialization errors in practice with a much lower annotation burden.

\nullaway's initialization checking was inspired by the checking performed in Eradicate, which also supports the \initannot annotation.  Compared with Eradicate, there are two main differences in how \nullaway checks for initialization. First, \nullaway only considers initialization from callees that are always invoked (see \autoref{sec:field-init}).  In contrast, Eradicate considers initialization performed in all (private or final) constructor callees, even those invoked conditionally, which is less sound.  E.g., if \autoref{li:init-helper-invoke} were written as \code{if (cond()) helper();}, Eradicate would still treat fields assigned in \code{helper} as initialized.  Second, Eradicate does not have any checking for use before initialization (\autoref{sec:use-before-init}).

\begin{sloppypar}
Note that usage of \initannot can be dangerous, as \nullaway does not check that such methods are invoked before others.  In the \uber code base most usage of \initannot is via overriding of well-known framework methods like \code{Activity.onCreate}.  When developers introduce new usage of \initannot, our code review system automatically adds a comment to warn about the risks.
\end{sloppypar}

\section{Purity Assumptions}\label{sec:side-effects}

\begin{figure}
\begin{lstlisting}
class FooHolder {
  @Nullable Object foo;
  public @Nullable Object getFoo() { return this.foo; }
  public void setFoo(@Nullable Object foo) {
    this.foo = foo;
  } 
  public @Nullable Object getFooOrNull() { /*@ \label{li:getfooornull-decl} @*/
    return randInt() > 10 ? null : this.foo;
  }
}
\end{lstlisting}
\caption{An example to illustrate \nullaway's purity handling.}
\label{fig:pure-code-ex}
\end{figure}

\nullaway reduces warnings (unsoundly) by assuming all methods are \emph{pure}, i.e., both side-effect-free and deterministic. \autoref{fig:pure-code-ex} gives a simple example of a class \code{FooHolder} that has a \code{foo} field with a getter and setter.  \nullaway's flow-sensitive type inference assumes method calls are side-effect-free, so it will (erroneously) not report a warning on this code:
\begin{lstlisting}
FooHolder f = ...;
if (f.foo != null) { /*@ \label{li:foo-null-check} @*/
  f.setFoo(null);
  f.foo.toString(); // NPE! /*@ \label{li:foo-null-deref} @*/
}
\end{lstlisting}
\nullaway ignores the effect of the \code{setFoo()} call and assumes \code{f.foo} remains non-null at \autoref{li:foo-null-deref}, based on the null check at \autoref{li:foo-null-check}.  
Additionally, \nullaway assumes all methods are deterministic, in order to refine nullability of ``getter'' return values during type inference.  The following code may throw an NPE:
\begin{lstlisting}
FooHolder f = ...;
if (f.getFooOrNull() != null) {
  f.getFooOrNull().toString();
}
\end{lstlisting}
The issue is that \code{getFooOrNull()} (defined at \autoref{li:getfooornull-decl} in \autoref{fig:pure-code-ex}) is non-deterministic: given the same parameters, it may return \code{null} in some calls but not others.  \nullaway ignores this possibility and refines the nullability of \code{getFooOrNull()}'s return to be \nonnull under the condition, and hence emits no warning.

\mypara{Discussion} In practice, we have not observed any NPEs in the field due to method side effects.  In the \uber code base most data-holding classes are immutable, precluding such errors.  Also, usually a null check is quickly followed by a dereference (with no intervening code), a safe pattern even with mutable types.  We have also not observed non-determinism to cause soundness issues for \nullaway in practice.  

By default, \cfnullness soundly assumes that methods may be impure.  While this catches more bugs, on the \uber code base this would lead to a large number of false warnings.  \cfnullness has an option to assume methods are side-effect free, but no option as of yet to assume determinism.  Previous work~\cite{DBLP:conf/cc/Pearce11,DBLP:conf/ccs/FinifterMSW08} has studied automatic verification of method purity for Java; it would be interesting future work extend \nullaway to verify these properties efficiently.


\section{Handling Unannotated Code}\label{sec:unannotated}

This section details how \nullaway handles interactions with unannotated, unchecked code, typically written by a third-party.  Since modern Java programs often use many third-party libraries without nullability annotations, these interactions arise frequently in real-world code.  By default, \nullaway uses an unsound, \emph{optimistic} handling of interactions with unannotated code, sacrificing some safety to enhance tool usability.


Assume that code in a program has been partitioned into \emph{checked} code, which has proper nullability annotations checked by \nullaway, and \emph{unannotated} code, which 
is lacking annotations and has not been checked.  (We shall detail how this partition is computed shortly.)  By default, \nullaway treats interactions between the checked and unannotated code \emph{optimistically}, i.e., it assumes that no errors will arise from the interaction.  In particular, this means:
\begin{itemize}
\item When checking a call to an unannotated method $m$, \nullaway assumes that $m$'s parameters are \nullable and that $m$'s return is \nonnull.
\item When checking an override of an unannotated method $m$ (see discussion of override checking in \autoref{sec:overview}), \nullaway assumes that $m$'s parameters are \nonnull and that $m$'s return is \nullable.
\end{itemize}
These assumptions are maximally permissive and ensure that no errors will be reported for interactions with unannotated code, a clearly unsound treatment.


Alternatives to optimistic handling of unannotated code yield too many false positives to be usable.  No handling of third-party code can prevent all NPEs, as there may be bugs within the third-party code independent of what values are passed to API methods.  A maximally-safe handling of interactions with third-party code would be \emph{pessimistic}, making the exact opposite assumptions from optimistic checking (e.g., all return values would be treated as \nullable).  But, these conservative assumptions lead to a huge number of false warnings.  By default, \cfnullness handles third-party libraries the same way as first-party code: any parameter or return missing an annotation is assumed to be \nonnull.  These assumptions also lead to a large number of false warnings (see \autoref{sec:eval}).

Granullar~\cite{DBLP:conf/cc/BrotherstonDL17} inserts runtime checks at the unannotated code boundary to guarantee soundness of checked code annotations.  We did not investigate this approach due to potential runtime overhead and the riskiness of shipping modified code.  

\begin{figure}
\begin{adjustbox}{width=\columnwidth,center}
\begin{tikzpicture} [node distance = 1cm, auto]
    \footnotesize
    \node [cloud, text width=10em] (foo) {\lstinline[mathescape]{a.b.C.foo($\colorbox{black!20}{\makebox(3,3){\textcolor{black!80}{?}}}$ Object o)}};
    \node [block, below of=foo] (annotated_packages) {\code{a.b} in \code{-XepOpt:...:AnnotatedPackages} regex?};
    \node [block, below of=annotated_packages] (unannotated_subpackages) {\code{a.b} in \code{-XepOpt:...:UnannotatedSubPackages} regex?};
    \node [block, below of=unannotated_subpackages] (unannotated_class) {\code{a.b.C} in \code{-XepOpt:...:UnannotatedClasses}?};
    \node [block, below of=unannotated_class] (generated) {Is class \code{C} annotated \code{@Generated} and \code{-XepOpt:...:TreatGeneratedAsUnannotated} set?};
    \node [block, below of=generated] (models) {Does \code{a.b.C.foo} have a library model for its first parameter?};
    \node [block, below of=models] (restrictive) {\code{AcknowledgeRestrictiveAnnotations} set and bytecode has \nonnull};
    \node [block, below of=restrictive] (jarinfer) {\code{JarInferEnabled} set and bytecode analyzed as requiring \nonnull};
    \node [cloud, below of=jarinfer, yshift=0.2cm] (from_model) {\code{From model}};
    \node [cloud, right of=from_model, xshift=0.5cm] (nonnull) {\nonnull};
    \node [cloud, left of=from_model, xshift=-0.5cm] (nullable) {\nullable};
    \path [line] (foo) -- (annotated_packages);
    \path [line] (annotated_packages) -- node {yes} (unannotated_subpackages);
    \path [line] (annotated_packages.west) -| node[above left,pos=0.1] {no} +(-1,-1) |- (models.west);
    \path [line] (unannotated_subpackages) -- node{no} (unannotated_class);
    \path [line] (unannotated_subpackages.west) -| node[above left,pos=0.1] {yes} +(-1,-1) |- (models.west);
    \path [line] (unannotated_class) -- node{no} (generated);
    \path [line] (unannotated_class.west) -| node[above left,pos=0.1] {yes} +(-1,-1) |- (models.west);
    \path [line] (generated.west) -| node[above left,pos=0.1] {yes} +(-1,-1) |- (models.west);
    \path [line] (generated.east) -| node[above right,pos=0.1] {no} +(1,-1) |- (nonnull.east);

    \path [line] (models) -- node{no} (restrictive);
    \path [line] (models.east) -| node[above right,pos=0.1] {yes} +(0.7,-1) |- +(-2.5,-3) -- (from_model.north);
    \path [line] (restrictive) -- node{no} (jarinfer);
    \path [line] (restrictive.east) -| node[above right,pos=0] {yes} +(1,-1) |- (nonnull.east);
    \path [line] (jarinfer) -- node[below right, pos=0.2]{no} (nullable);
    \path [line] (jarinfer) -- node[below left, pos=0.2]{yes} (nonnull);
    \path [line, dotted] (from_model) -- (nullable);
    \path [line, dotted] (from_model) -- (nonnull);
\end{tikzpicture}
\end{adjustbox}
\caption{Flowchart for \nullaway's treatment of unannotated code.}
\label{fig:unannotated-code-options}
\end{figure}
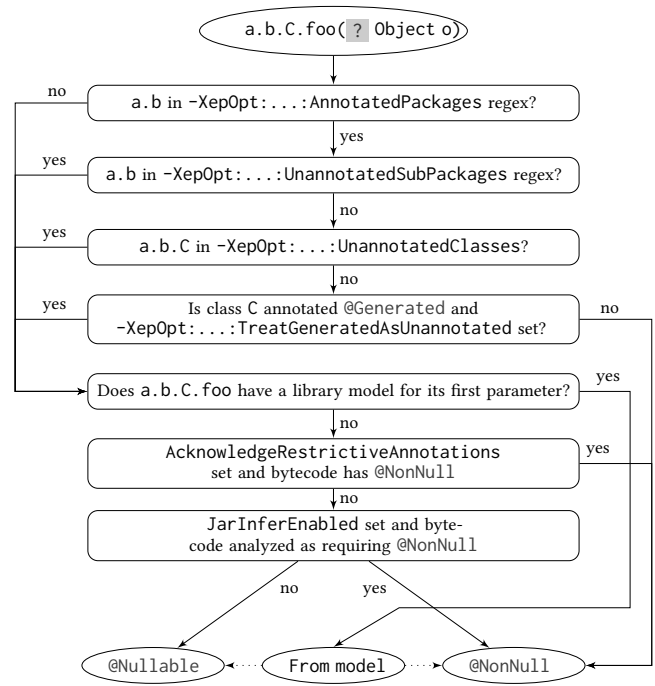

\nullaway has a highly-configurable system for specifying which code is unannotated and how optimistically it is handled.  At the highest level, annotated and unannotated code is partitioned based on its Java package, \emph{not} whether the code is first-party or third-party.  This system provides a high degree of flexibility when adopting \nullaway; source packages can be treated as unannotated for gradual adoption, while third-party packages can be treated as annotated if they have proper annotations present.

\begin{sloppypar}
\autoref{fig:unannotated-code-options} presents a flow chart showing how \nullaway determines the nullability of the first parameter to a hypothetical method \code{a.b.C.foo(Object o)} (represented by the missing annotation placeholder $\colorbox{black!20}{\makebox(3,3){\textcolor{black!80}{?}}}$). The first four steps seek to determine whether the code is annotated or unannotated.  The method is treated as annotated if (1) the package name matches the \code{AnnotatedPackages} regex, (2) it does not match the \code{UnannotatedSubPackages} regex, (3) the class name is not blacklisted in \code{UnannotatedClasses}, and (4) the class is not annotated as \code{@Generated} with the option \code{TreatGeneratedAsUnannotated} set. In this case, the nullability of \code{o} is assumed to be \nonnull.
\end{sloppypar}

Otherwise, the code is unannotated.  \nullaway then checks if there is a manually-written library model giving an annotation for the method parameter; if so, that annotation is used. \nullaway ships with 95 such models, one per method and parameter position pair. These mostly cover common methods from the JDK and Android SDK. \nullaway can load additional custom library models, but none of the open-source apps in our evaluation required it.

If the \code{AcknowledgeRestrictiveAnnotations} option is set, \nullaway looks for explicit annotations within unannotated code, using them if they are more restrictive than its default assumptions.  This allows \nullaway to opportunistically take advantage of explicitly-annotated third-party code, without forcing its default assumptions for checked code onto unannotated methods.  Here, if \code{foo}'s parameter had an explicit \nonnull annotation, it would be used.

Finally, \nullaway can leverage models automatically generated by JarInfer, a separate analysis we built for doing basic type inference on bytecode.  For example, if a method unconditionally dereferences its parameter, JarInfer infers that the parameter should be \nonnull.  While JarInfer still only performs intra-procedural analysis on library entrypoints, we have found it useful at \uber for catching additional issues in interactions with libraries.  A full description of JarInfer is outside the scope of this paper; we plan to extend it with greater functionality and present it in future work.  None of the open-source apps in our evaluation use JarInfer.


\section{Other Features}\label{sec:other-features}

In this section we detail \nullaway's handling of other Java features relevant to nullability, and describe some additional \nullaway features that reduce false warnings.

\mypara{Polymorphism} \nullaway does not yet support polymorphic nullability for generic types.  Consider the following \code{Pair} type:
\begin{lstlisting}
class Pair<T,U> {
  public T first; public U second;
}
\end{lstlisting}
\cfnullness allows for different uses of \code{Pair} to vary in nullability via type-use annotations on generic type parameters~\cite{PapiACPE2008}.  E.g., one can write \code{Pair<@Nullable String, String>} for \code{Pair}s where the first element may be null.  In contrast, \nullaway treats generic types like any other; for \code{Pair}, it assumes both fields have the \nonnull default.  To allow \code{null} as a value, the fields themselves would need to be annotated \nullable.  Type-use annotations on the generic type parameters are ignored.  This treatment is sound but could lead to undesirable code duplication; e.g., one may need to write a nearly-identical \code{FirstNullablePair}, \code{SecondNullablePair}, etc.

We have found this lack of support for type polymorphism to be only a minor annoyance thus far.  A big mitigating factor is that most generic type usages in the \uber codebase are of \code{Collection} data structures, and it is a best practice to simply avoid storing \code{null} in such types~\cite{guava-collections-null}.  However, we do see a need to eventually support type polymorphism, for cases like the above \code{Pair} type and also functional interface types like those in the \code{java.util.function} package.  We plan to add support in future work, but doing so without compromising on build time overhead may require care.

\mypara{Arrays} \nullaway unsoundly assumes that arrays do not contain \code{null} values.  In contrast, \cfnullness uses type-use annotations to reason about nullability of array contents; e.g., the type of an array of possibly-null \code{String}s is written \code{@Nullable String []}.  (Note that \cfnullness does not soundly check array initialization by default~\cite[\S 3.3.4]{cf-manual}.)  In the \uber code base, arrays of references are rarely used; \code{Collection}s are used instead.  For more array-intensive code, this \nullaway unsoundness could have a greater impact.

\mypara{Lambdas} In Java, parameters to lambda expressions do not require explicit types, instead their parameter and return types are usually inferred from those of the single method in the corresponding functional interface.  Analogous to this, \nullaway uses the annotations of that same functional interface method to determine the nullability of the parameters and return value of the lambda.

\mypara{Handlers} \nullaway provides an internal extension mechanism called handlers. A handler implements a known interface to interpose itself at specific plug-in points during the analysis process and alter the nullability information available to \nullaway. The following two features are implemented as handlers.

\mypara{Pre- and post-conditions} \nullaway supports simple pre- and post-condition specifications via partial support for \code{@Contract} annotations~\cite{jb-contract}.  Here are some example usages of \code{@Contract} supported by \nullaway:
\begin{lstlisting}
public class NullnessHelper {
  $$@Contract("null -> false")
  static boolean isNonNull(@Nullable Object o) {
    return o != null;
  }
  $$@Contract("null -> fail")
  static void assertNonNull(@Nullable Object o) {
    if (o == null) throw new Error();
  }
  $$@Contract("!null -> !null")
  static @Nullable Object id(@Nullable Object o) {
    return o;
  }
}
\end{lstlisting}
The \code{@Contract} annotations document that \code{isNonNull} returns \code{false} when passed \code{null}, and that \code{assertNonNull} fails when passed \code{null}.  The annotation on \code{id} indicates that if passed a non-null value, a non-null value is returned, yielding some support for parametric polymorphism (like \code{@PolyNull} in \cfnullness~\cite{PapiACPE2008}).  Currently, \nullaway trusts \code{@Contract} annotations, but we plan to add checking for them soon.  

\mypara{Streams} While \nullaway's general type inference/refinement is strictly intra-procedural, handlers can propagate nullability information inter-procedurally for a few well-understood APIs. At \uber we do this mostly for stream APIs like RxJava \cite{lib-rxjava}. Consider the following code using a common filter and map pattern:
\begin{lstlisting}
public class Baz { @Nullable Object f; ... }
public class StreamExample {
	public void foo(Observable<Baz> o) {
	  o.filter(v -> v.f != null)
	   .map(v -> v.f.toString());
	}
}
\end{lstlisting}
In the above example, there are three separate procedures: \code{foo}, and the two lambdas passed to the \code{filter} and \code{map} method calls. The lambda for \code{filter} will filter out any value in the stream for which \code{v.f} is \code{null}.  For the lambda inside \code{map}, \nullaway's usual analysis would emit an error, due to the call of \code{toString()} on the \nullable field \code{v.f}.  But this code is safe, as objects with a null \code{f} field were already filtered out. \nullaway includes a handler that analyzes the exit statements of every lambda passed to \code{Observable.filter}, and the entry statement of every lambda passed to \code{Observable.map}. If the \code{map} call is chained immediately after the \code{filter} call (as in the previous example), this handler propagates the nullability information for the parameter of the \code{filter} lambda on exit (conditioned on the return value being \code{true}) to the parameter of the \code{map} lambda.  In the example above, when the \code{filter} lambda returns \code{true}, \code{v.f} must be \nonnull.  This fact gets preserved at the entry of the \code{map} lambda, and hence \nullaway no longer reports a warning at the \code{toString()} call.  This heuristic handles common cases observed in the \uber code base and reduces the need for warning suppressions, without introducing any new unsoundness.

%% file: sections/implementation.tex

\section{Implementation and Deployment}\label{sec:implementation}


\nullaway is built as a plugin to the Error Prone framework for compile-time bug finding~\cite{DBLP:conf/scam/AftandilianSPK12,tool-ep}.  Error Prone is carefully designed to ensure its checks run with low overhead, enabling the checking to run on every build of a project.  Checks leverage the parsing and type checking already done by the Java compiler, thereby avoiding redundant work.  Further, Error Prone interleaves running all checks in a \emph{single pass} over the AST of each source file, a more efficient architecture than doing a separate AST traversal per check.

The core of \nullaway primarily adheres to the single-pass architecture encouraged by Error Prone.  Some additional AST traversal is required to collect class-wide information up front like which fields are \nonnull, to facilitate initialization checking (\autoref{sec:init-checking}).  To perform flow-sensitive type inference, \nullaway uses the standalone Checker Framework dataflow library~\cite{cf-dataflow-manual} to construct control-flow graphs (CFGs) and run dataflow analysis.  CFG construction and dataflow analysis are by far the most costly operations performed by \nullaway.  The tool employs caching to ensure that dataflow analysis is run at most once per method and reused across standard type checking and initialization checking.  We profile \nullaway regularly to ensure that performance has not regressed.

\nullaway has been deployed at \uber for nearly two years. For Android code, \nullaway runs on every compilation (both locally and during continuous integration), blocking any change that triggers a nullness warning from merging to the main master branch.
Test code and third-party libraries are treated as unannotated, with both restrictive annotations and JarInfer enabled (see \autoref{sec:unannotated}). 

%% file: sections/evaluation.tex

\section{Evaluation}	\label{sec:eval}


We evaluated \nullaway on a diverse suite of open-source Java programs and via a long-term deployment on \uber's Android codebase.  The evaluation targeted the following research questions:
\begin{description}[leftmargin=*]
\item[RQ1:] What is the annotation burden for using \nullaway?
\item[RQ2:] How much build-time overhead does \nullaway introduce, and how does this overhead compare to previous tools?
\item[RQ3:] Do \nullaway's unsound assumptions for checked code lead to missed NPE bugs?
\item[RQ4:] Compared to checking with \cfnullness, how much do each of \nullaway's unsound assumptions contribute to the reduction in warning count?
\end{description}

\subsection{Experimental Setup} \label{sec:eval-setup}

To evaluate \nullaway's effectiveness on open-source software, we gathered a diverse suite of benchmark Java projects from GitHub that already utilize \nullaway.  We searched for all projects integrating \nullaway via the Gradle build system, and then included all non-duplicate projects that we could successfully build, excluding small demo projects.  The projects vary in size and domain---they include widely-used Java, Android, and RxJava libraries, as well as some Android and Spring applications.

\begin{table}
\footnotesize
\centering
\caption{\textnormal{Benchmark Java Projects}\vspace{-1em}}
{\renewcommand{\arraystretch}{1.2}
\begin{tabular}{|c|l||r|r|r|}
\hline
\multicolumn{2}{|c||}{\multirow{2}{*}{Benchmark Name}} & \multirow{2}{*}{KLoC} & \multicolumn{2}{c|}{Annotations per KLoC} \\\cline{4-5}
\multicolumn{2}{|c||}{} & & Nullability & Suppression \\\hline\hline
\multicolumn{2}{|l||}{\uber repository} &    \~3.3 MLoc & 11.82\hspace{1em} & 0.15\hspace{1em} \\\hline\hline
\multirow{24}{*}{\rotatebox[origin=c]{90}{Open Source Projects}}
& \multicolumn{4}{l|}{Build Tools} \\\cline{2-5}
& \hspace{1em} \texttt{okbuck}               &  8.96 & 13.06\hspace{1em} &  0.67\hspace{1em} \\\cline{2-5}
& \multicolumn{4}{l|}{Libraries - Android} \\\cline{2-5}
& \hspace{1em} \texttt{butterknife}          & 15.55 &  3.47\hspace{1em} &  0.06\hspace{1em} \\\cline{2-5}
& \hspace{1em} \texttt{picasso}              &  9.56 & 11.61\hspace{1em} &  0.21\hspace{1em} \\\cline{2-5}
& \hspace{1em} \texttt{RIBs}                 &  9.43 & 32.45\hspace{1em} &  0.64\hspace{1em} \\\cline{2-5}
& \hspace{1em} \texttt{FloatingSpeedDial}    &  2.21 & 28.51\hspace{1em} &  0.00\hspace{1em} \\\cline{2-5}
& \hspace{1em} \texttt{uLeak}                &  1.38 &  3.62\hspace{1em} &  2.90\hspace{1em} \\\cline{2-5}
& \multicolumn{4}{l|}{Libraries - RxJava} \\\cline{2-5}
& \hspace{1em} \texttt{AutoDispose}          &  8.27 &  5.32\hspace{1em} &  0.48\hspace{1em} \\\cline{2-5}
& \hspace{1em} \texttt{ReactiveNetwork}      &  2.16 &  0.00\hspace{1em} &  6.02\hspace{1em} \\\cline{2-5}
& \hspace{1em} \texttt{keyvaluestore}        &  1.40 & 11.43\hspace{1em} &  0.00\hspace{1em} \\\cline{2-5}
& \multicolumn{4}{l|}{Libraries - Other} \\\cline{2-5}
& \hspace{1em} \texttt{caffeine}             & 51.72 &  5.84\hspace{1em} & 11.06\hspace{1em} \\\cline{2-5}
& \hspace{1em} \texttt{jib}                  & 27.14 & 15.59\hspace{1em} &  0.04\hspace{1em} \\\cline{2-5}
& \hspace{1em} \texttt{skaffold-tools}       &  1.29 &  5.43\hspace{1em} &  0.00\hspace{1em} \\\cline{2-5}
& \hspace{1em} \texttt{filesystem-generator} &  0.14 &  0.00\hspace{1em} &  0.00\hspace{1em} \\\cline{2-5}
& \multicolumn{4}{l|}{Apps - Android} \\\cline{2-5}
& \hspace{1em} \texttt{QRContact}            &  9.99 & 11.31\hspace{1em} &  0.20\hspace{1em} \\\cline{2-5}
& \hspace{1em} \texttt{test-ribs}            &  6.29 & 24.64\hspace{1em} &  0.95\hspace{1em} \\\cline{2-5}
& \hspace{1em} \texttt{ColdSnap}             &  5.13 & 24.37\hspace{1em} &  0.00\hspace{1em} \\\cline{2-5}
& \hspace{1em} \texttt{OANDAFX}              &  0.99 & 46.46\hspace{1em} &  0.00\hspace{1em} \\\cline{2-5}
& \multicolumn{4}{l|}{Apps - Spring} \\\cline{2-5}
& \hspace{1em} \texttt{meal-planner}         &  2.62 &  1.15\hspace{1em} &  0.00\hspace{1em} \\\cline{2-5}
& \multicolumn{4}{l|}{Average (Open Source Projects)} \\\cline{2-5}
& &  9.12 &  13.57\hspace{1em} &  1.29\hspace{1em} \\\hline
\end{tabular}}
\label{tab:benchmarks}
\end{table}

\autoref{tab:benchmarks} summarizes the details of our benchmark suite, including the internal code base at \uber.  Regarding RQ1, \nullaway's annotation burden is quite reasonable, with $11.82$ nullability-related annotations per KLoc on the \uber code base, and $13.57$ such annotations per KLoc on the open-source benchmarks.  As observed in previous work~\cite{PapiACPE2008}, using \nonnull as the default both decreases the annotation burden and encourages better coding style.

Our experimental harness ran as follows. First, we ensured that all projects built without any warnings using \nullaway 0.6.4; the numbers in \autoref{tab:benchmarks} include additional annotations required for a few cases. The harness captured the compiler arguments for each build target in each project based on Gradle's verbose output.  Then it modified the arguments as needed to run each build with \nullaway, \cfnullness~\cite{tool-cf}, and Eradicate~\cite{tool-eradicate}.  We ran all tools in their default configuration;\footnote{Note that in its default configuration, \cfnullness employs unsound assumptions around array initialization and handling of class files. See \url{https://checkerframework.org/manual/\#nullness-arrays} and \url{https://checkerframework.org/manual/\#defaults-classfile} for details.  \cfnullness is still always more strict than \nullaway.} for \nullaway the only preserved setting was the set of annotated packages.

To answer RQ2, we measured the overhead of each run against the time to run the standard Java compiler with no nullness checking.  All experiments on the open-source apps were run on a single core of an Intel Xeon E5-2620 processor with 16GB RAM running Linux 4.4, and Java JDK 8. We used \cfnullness v.2.8.1 and Infer v.0.15.0.
Due to the size and complexity of \uber's build environment, we did not attempt to run other tools there; we still measure \nullaway's overhead compared to compilation with it disabled.

%
%
%

To answer RQ3, we studied all NPEs present in production crash data on \uber's applications over a period of 30 days, looking for cases where \nullaway's unsound assumptions on checked code led to crashes.  \uber's crash reporting infrastructure de-duplicates crash instances with the same stack trace. For the 30-day period we studied, there were 100 distinct stack traces involving NPEs. This includes crashes in both internal and production versions of the app, for all versions in use during the time period, possibly including months-old versions.  We included all crashes to get the broadest dataset of NPEs in code that had passed \nullaway's checks.

Additionally, for the open-source benchmarks, we manually inspected a random subset of the additional warnings given by \cfnullness as compared to \nullaway.  As further evidence for RQ3, we checked if the warnings corresponded to real bugs.  For RQ4, we categorized each warning based on which unsound assumption led to its omission by \nullaway.

Regarding the precision of Eradicate as compared to \nullaway, we found that doing a proper comparison would be non-trivial.  Eradicate does not yet support recent Java 8 language features like lambdas and method references, and evaluating the full impact of this difference on Eradicate's false negative rate would require significant additional experiments beyond the scope of this paper.

\mypara{Data Availability} \nullaway and the scripts required to run our evaluation on the open-source benchmarks are publicly available \cite{tool-nullaway, na-eval-repo}. We have also provided our raw experimental data as supplementary material \cite{supp-data}.

%
%

%

\subsection{Compile Time Overheads} \label{sec:eval-performance}

\begin{figure*}[t]
\begin{center}
\includegraphics[width=2.1\columnwidth]{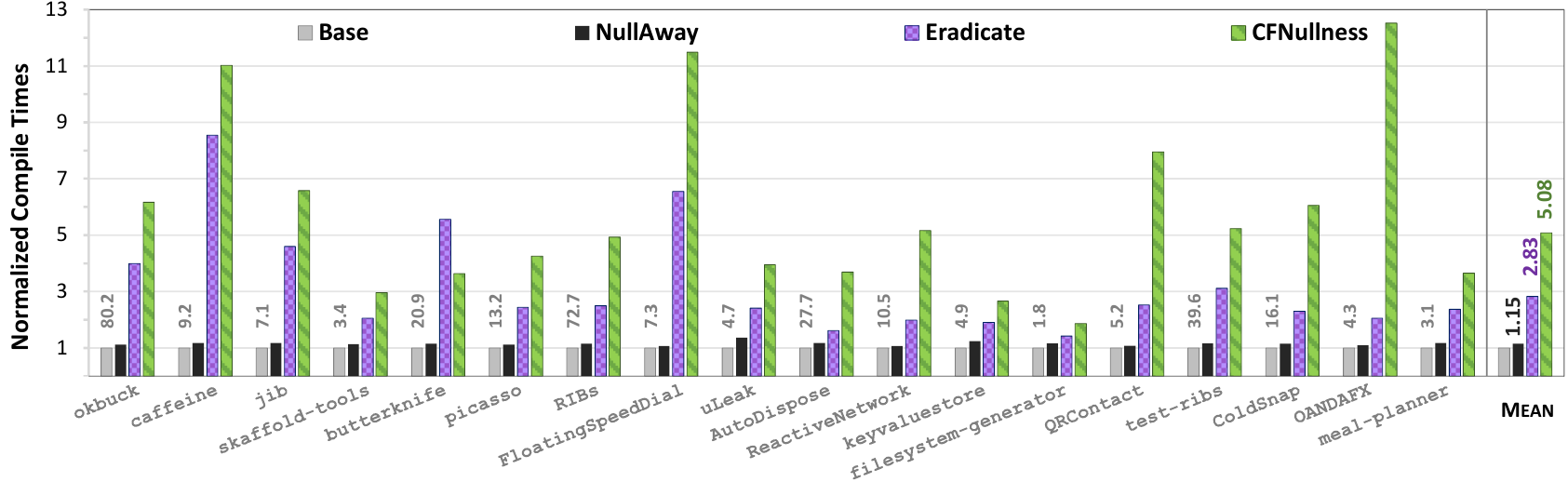}\vspace{-0.75em}
\caption{Build-time overheads of \nullaway, \cfnullness, and Eradicate.  \textnormal{Compile times are normalized to the `Base' compile times without nullness checking.  Absolute times for Base compiles are labeled above the bars (in seconds).}\vspace{-1em}}
\label{fig:eval-perf}
\end{center}
\end{figure*}



\begin{sloppypar}
\autoref{fig:eval-perf} shows the build-time overheads of the tested nullness-checking tools as compared to compilation without nullness checking.  On average, \nullaway's build times are only $1.15\times$ those of standard builds, compared to $2.8\times$ for Eradicate and $5.1\times$ for \cfnullness.  In fact, \nullaway's highest observed overhead ($1.36\times$ for \texttt{uLeak}) is close to Eradicate's lowest ($1.43\times$ for \texttt{filesystem-generator}). Our supplementary materials \cite{supp-data} give full data on absolute compilation times and overheads for all runs.
\end{sloppypar}

Though \cfnullness also runs as part of the Java compiler, we conjecture that its overheads are significantly higher than \nullaway's due to its greater sophistication and focus on ease-of-use rather than performance.\footnote{This discussion is based on personal communication with \cfnullness developers.}  \cfnullness does significantly more complex checking than \nullaway, including full checking and inference of generic types and support for tracking map keys~\cite[\S4]{cf-manual}.  Also, the Checker Framework has been designed to make writing new checkers easy, with much implementation shared in a common base type checker.  This architecture does not yet minimize costly operations like AST passes and dataflow analysis runs.

We note that developers often perform incremental builds, in which only modified source files or targets and their dependencies are recompiled.  Such builds are typically much faster than doing a clean rebuild of an entire project, so the overhead of nullness checking tools will consume less absolute time.  Nevertheless, it is our experience that even with incremental builds, the overhead levels of Eradicate and \cfnullness would still be a significant negative impact on developer productivity if run on every build.

For the \nullaway deployment at \uber, measuring overhead is difficult due to use of modular and parallel builds, network-based caching, compilation daemons, and other annotation processors and analyses.  As an estimate of overhead, we ran five builds of the entire monorepo with all forms of caching disabled, comparing our standard build with a build where \nullaway is disabled, and we observed a $1.22\times$ overhead on average.

To summarize: \nullaway has \emph{significantly lower compile time overhead} than previous tools, and hence can be enabled \emph{for every build} on large codebases.  By running on local builds with low overhead, \nullaway helps developers catch potential NPEs early, improving code quality and productivity.

\subsection{\nullaway and NPEs at \uber} \label{sec:eval-uber}

\begin{table}
\caption{\textnormal{Classification of NPEs from the \uber deployment}\vspace{-1em}}
\begin{adjustbox}{width=\columnwidth,center}
{\renewcommand{\arraystretch}{1.2}
\begin{tabular}{|l|l|c|}
\hline
Category & Sub-category & Count \\
\hline
\hline
\multirow{4}{*}{Unannotated library code} & Android SDK / JDK & 38 \\
& Other Third-Party & 16 \\
& First Party Libs & 10 \\
\cline{2-3}
& Total & \textbf{64} \\
\hline
\hline
\multirow{3}{*}{Manual Suppressions} & Precondition and assertion-like methods & 14 \\
& \code{@SuppressWarnings} annotations & 3 \\
\cline{2-3}
& Total & \textbf{17} \\
\hline
\hline
\multirow{5}{*}{Post-checking issues} & Reflection & 10 \\
& Instrumentation & 3 \\
& Annotation Processor Misconfiguration & 2 \\
& Code stripping & 2 \\
\cline{2-3}
& Total & \textbf{17} \\
\hline
\hline
Other & & \textbf{2} \\
\hline
\hline
Total & & \textbf{100} \\
\hline
\end{tabular}}
\end{adjustbox}
\label{tab:npes}
\end{table}

To aid in answering RQ3, \autoref{tab:npes} gives a categorization of NPEs observed in \uber Android apps over a 30-day period.  NPEs were de-duplicated based on the full stack-trace of the exception, meaning that the same root cause can be counted multiple times (e.g., for a bug in a shared library used by multiple apps).

Ideally, we would like to compare the rate of NPEs at \uber before and after the introduction of static nullness checking. However, between a previous deployment of Eradicate and \nullaway itself, \uber's code base has been running some null checker for over 2 years, and we do not have data to do this comparison. We do note that the documented motivation for adopting these tools was the prevalence of NPEs as a category of crashes. Today, NPEs comprise less than $5\%$ of observed crashes for our Android apps.

The most common type of NPEs missed by our tool ($64\%$) are those involving unannotated third-party code.  This case includes crashes within unannotated code and cases where unannotated methods returned \code{null} (\nullaway optimistically assumes a non-null return value). Note that these cases were not necessarily bugs in the libraries; they could be a misunderstood API contract due to lack of nullness annotations.  $38\%$ of the crash stacks involved Android framework libraries or (rarely) the JDK classes, $16\%$ involved other third-party libraries, and $10\%$ involved \uber libraries outside the Android monorepo that do not build with \nullaway.

\begin{sloppypar}
The next broad category ($17\%$) were manual suppressions of \nullaway warnings.  Only $3\%$ were explicit \code{@SuppressWarnings} annotations, while $14\%$ were calls to methods that perform a runtime check and crash when passed \code{null} (e.g. Guava's \code{Preconditions.checkNotNull}~\cite{lib-guava}).  Such calls are warning suppressions since \nullaway treats the parameter as \nonnull after the call without proving that the call cannot fail.  In most cases, these suppressions were either introduced to speed initial \nullaway adoption and never fixed, or cases where developers mistakenly concluded non-nullness was enforced by some invariant outside the type system.
\end{sloppypar}

An equally large category of crashes ($17\%$), involved cases in which code was transformed in some way after checking.  The most prevalent case was due to an object serialization framework unexpectedly writing \code{null} into a \nonnull field via reflection ($10\%$).  We also observed issues due to dynamic instrumentation breaking the assumptions made by the compiler ($3\%$), misconfiguration of annotation processors resulting on incorrect or incomplete generated code ($2\%$), and erroneous code stripping during the final steps of the build process ($2\%$).

Finally, two uncategorized cases remain: one was an explicitly-thrown \code{NullPointerException} unrelated to nullability, and one could not be triaged correctly based on available data.

Critically, not a single NPE in this dataset was due to any of \nullaway's unsound assumptions for fully-checked code; they only related to handling of unannotated third-party code or code transformations (e.g. reflection) that are out of scope for nearly all static analysis tools.  In other words, for this dataset, \nullaway achieved its goal of \emph{no false negatives in practice for checked code}.

\subsection{Additional \cfnullness Warnings} \label{sec:eval-cfnullness-warnings} 
\begin{table*}[htb]
\begin{threeparttable}
\footnotesize
\centering
\caption{\textnormal{Additional nullness warnings with \cfnullness}\vspace{-1em}}
\begin{tabular}{lrcccccc}\toprule
\mrCell{3}{0}{Benchmark Name} & \mrCell{3}{0.5}{\hspace{-3.5em}\# Warnings} & \mcCell{2}{3.5}{Initialization Checking} & \mcCell{2}{2.5}{Unannotated Code} & \mrCell{3}{1.5}{Arrays / Purity} & \mrCell{3}{1}{Suppressed}\\\cline{3-6}
& & \mrCell{2}{1.15}{Helper methods} & \mrCell{2}{1.5}{\initannot methods} & \mrCell{2}{1.5}{Optimistic defaults} & \mrCell{2}{0.85}{Library models} & &\\
& & & & & & &\\\toprule
\texttt{okbuck}               &   17 &   & 1 & 6 & 3 &   &   \\
\texttt{caffeine}             &   47 &   & 3 &   & 5 & 2 &   \\
\texttt{jib}                  &  141 & 2 & 1 & 4 & 1 & 2 &   \\
\texttt{skaffold-tools}       &   12 &   & 1 &   & 6 & 3 &   \\
\texttt{butterknife}          &   58 & 2 & 1 & 1 & 6 &   &   \\
\texttt{picasso}              &   31 &   & 2 & 4 & 4 &   &   \\
\texttt{RIBs}                 &   22 &   & 4 & 3 & 2 & 1 &   \\
\texttt{FloatingSpeedDial}    &    5 & 1 & 2 & 2 &   &   &   \\
\texttt{AutoDispose}          &   21 & 3 & 1 & 5 &   &   & 1 \\
\texttt{ReactiveNetwork}      &    6 &   & 1 & 2 &   &   & 3 \\
\texttt{keyvaluestore}        &    7 &   & 1 & 6 &   &   &   \\
\texttt{QRContact}            &    4 &   &   & 4 &   &   &   \\
\texttt{test-ribs}            &   22 & 1 & 2 & 4 & 2 &   & 1 \\
\texttt{meal-planner}         &   11 & 1 &   & 5 & 4 &   &   \\
\bottomrule\end{tabular}
\label{tab:warnings}
\end{threeparttable}
\end{table*}

We also inspected additional warnings emitted by \cfnullness for build targets in the open-source benchmarks that passed \nullaway with no warnings.  As shown in \autoref{tab:warnings}, \cfnullness raised a total of 404 additional warnings for 14 out of our 18 benchmark projects. We manually inspected 10 randomly-sampled warnings per benchmark, or all if there were fewer than 10.  For RQ3, we determined if they were true bugs missed by \nullaway (i.e., the code could cause an NPE at runtime).  For RQ4, we determined which unsound assumption caused \nullaway to not emit the warning.

Among the 122 randomly sampled warnings in \autoref{tab:warnings}, we found {\bf no true NPE issues} that were missed by \nullaway.   Going left to right in \autoref{tab:warnings}, 30 warnings ($25\%$) were due to stricter initialization checking in \cfnullness, specifically prohibiting helper method calls from constructors and lack of support for the \initannot annotation. 79 warnings ($65\%$) were in calls to code deemed unannotated by \nullaway.  \cfnullness treats third-party methods without annotations as having \nonnull parameters, whereas \nullaway's optimistic treatment assumes \nullable (see \autoref{sec:unannotated}).  We inspected these cases and did not find any true bugs, but as shown in \autoref{sec:eval-uber} such bugs are possible.  Also, some warnings were due to differences in built-in models of library code.  For the ``Arrays / Purity'' column, one case involved sound treatment of array accesses (\autoref{sec:other-features}), and the 7 other cases involved purity assumptions (\autoref{sec:side-effects}). Finally, 5 of the issues ($4\%$) would have generated \nullaway warnings but were explicitly suppressed (last column).

To conclude, we found that on the open-source benchmarks, \nullaway \emph{did not miss any real NPE issues}, despite being unsound.  Adding enough annotations to remove all \cfnullness warnings would require significant effort, so in these cases, use of \nullaway improved productivity without negative impact on safety.  Regarding the relative impact of \nullaway's various unsound assumptions on warning count, optimistic treatment of unannotated code reduced warnings the most, followed by initialization checking, with unsound handling of arrays and purity having a lesser impact.


\subsection{Threats to Validity}

The primary threat to the validity of our evaluation is our choice of benchmarks.  We chose open-source benchmarks that already used \nullaway, as getting unannotated code bases to compile with \nullaway would have required significant effort.  Also, \nullaway was built for and tuned to the style of the \uber code base.  Together these programs comprise millions of lines of code written by hundreds of developers, so we expect it to be representative of a large class of Java programs.  But it is possible that code bases that have adopted \nullaway are particularly suited to its checking style, and that on other types of programs (e.g., for scientific computing) \nullaway's unsoundness would lead to more real-world NPEs.

Regarding build-time overhead, a threat is that our experimental configuration may not reflect typical build configurations, which e.g. may use concurrency or a daemon process (see discussion in \autoref{sec:eval-performance}).  Our experience with the \uber deployment gives confidence that our measurements are reflective of typical builds.

%% file: sections/related.tex

\section{Related Work}\label{sec:related}

Other sections of the paper have compared \nullaway with \cfnullness~\cite{PapiACPE2008,DietlDEMS2011} and Eradicate~\cite{tool-eradicate}.  At \uber, Eradicate was used before \nullaway was built, and it succeeded in greatly reducing the number of NPEs observed in production.  The primary motivation for building \nullaway was to reduce overhead such that checking could be run on every build (Eradicate only ran during continuous integration).  Additionally, at the time \nullaway was initially built, Eradicate was not able to process nullability annotations from bytecode for which it had not also processed the corresponding source code, leading to many missed warnings; this Eradicate issue has since been fixed.

\nullaway's implementation was inspired by various partial nullability checks built into Error Prone~\cite{ep-nullderef-check,ep-retmissingnullable-check}.  These checks also leverage dataflow analysis for inference via the Checker Framework dataflow library~\cite{cf-dataflow-manual}, from which \nullaway's implementation borrows and extends.  While Error Prone's checks can catch a variety of issues, they are not as complete as \nullaway; e.g., they lack initialization checking (\autoref{sec:init-checking}) and method override checking.

As mentioned in \autoref{sec:init-checking}, checking of proper initialization has been the subject of a variety of previous work~\cite{DBLP:conf/oopsla/FahndrichL03, DBLP:conf/oopsla/SummersM11, DBLP:conf/popl/QiM09, DBLP:conf/oopsla/FahndrichX07}, as incorrect initialization jeopardizes reasoning about many other object properties (e.g., immutability~\cite{DBLP:conf/oopsla/ZibinPLAE10, DBLP:conf/oopsla/HuangMDE12}).  \nullaway is distinguished by initialization checking that is unsound, yet has prevented nearly all initialization errors in a multi-million line code base over many months, requiring fewer annotations than the sound approaches.

Many approaches have been proposed for preventing NPEs with static analysis~\cite{DBLP:conf/icse/NandaS09, DBLP:conf/issta/LoginovYCFRN08, DBLP:journals/jacm/CalcagnoDOY11, DBLP:conf/oopsla/MadhavanK11}.  These approaches work out of the box on a subject code base, without requiring developers to initially add annotations.  However, such approaches also have drawbacks.  Type-based approaches like \nullaway are modular and efficient, leveraging type annotations at method boundaries.  In contrast, most static analysis approaches are not modular, forcing an expensive global re-analysis at each code change.  The bi-abduction approach of Facebook Infer is modular~\cite{DBLP:journals/jacm/CalcagnoDOY11} but still significantly more complex and expensive than type checking.  Also, null pointer issues discovered via static analysis can be difficult for a developer to understand, as they can involve inter-procedural data flow.  In contrast, \nullaway errors always relate to a local violation of the typing discipline, and hence can be understood by only looking at surrounding code and related method type signatures.

\autoref{sec:eval} showed that though \nullaway is unsound, it has been able to prevent nearly all NPEs in production Android code at \uber.  The RacerD system for static data race detection~\cite{DBLP:journals/pacmpl/BlackshearGOS18} also aims for this threshold of few to no false negatives in practice, based on bugs observed in the field.  We anticipate that future work will continue this trend of static analysis and type system design based on preventing nearly all errors previously observed in testing and production, rather than aiming for strict soundness.

%% file: sections/conclusion.tex

\section{Conclusions and Future Work}\label{sec:conclusions}

We have presented \nullaway, a practical tool for type-based null safety of large-scale Java applications.    \nullaway has much lower build-time overhead than previous tools, enabling null checking on every build.  Further, \nullaway's checks have been carefully tuned to minimize false negatives in practice while imposing a reasonable annotation burden.  \nullaway runs on every build of millions of lines of code at \uber and has been adopted by many other open-source projects and companies.

As shown by \nullaway and other recent work~\cite{stein18rx}, developers are willing to use pluggable type systems, even with a moderate annotation burden, to prevent mobile application crashes, due to the difficulty of addressing such crashes in production.  The type systems need not be fully sound, as long as they provide reasonably strong safety guarantees in practice.  We believe that developing type systems with soundness tradeoffs similar to \nullaway for other common types of crashes (e.g., array indexing errors) is a fruitful area for future research.  Beyond investigating such systems, we are actively researching improved nullability type inference for libraries, to further increase the safety of \nullaway-checked code.

\mypara{Acknowledgements.}  We thank Werner Dietl and Michael D. Ernst for help with understanding \cfnullness and detailed feedback on drafts of this paper.

%% file: ms.bbl

\begin{thebibliography}{38}


\ifx \showCODEN    \undefined \def \showCODEN     #1{\unskip}     \fi
\ifx \showDOI      \undefined \def \showDOI       #1{#1}\fi
\ifx \showISBNx    \undefined \def \showISBNx     #1{\unskip}     \fi
\ifx \showISBNxiii \undefined \def \showISBNxiii  #1{\unskip}     \fi
\ifx \showISSN     \undefined \def \showISSN      #1{\unskip}     \fi
\ifx \showLCCN     \undefined \def \showLCCN      #1{\unskip}     \fi
\ifx \shownote     \undefined \def \shownote      #1{#1}          \fi
\ifx \showarticletitle \undefined \def \showarticletitle #1{#1}   \fi
\ifx \showURL      \undefined \def \showURL       {\relax}        \fi
\providecommand\bibfield[2]{#2}
\providecommand\bibinfo[2]{#2}
\providecommand\natexlab[1]{#1}
\providecommand\showeprint[2][]{arXiv:#2}

\bibitem[\protect\citeauthoryear{??}{cf-}{2019}]%
        {cf-manual}
 \bibinfo{year}{2019}\natexlab{}.
\newblock \bibinfo{title}{{Checker Framework Manual}}.
\newblock \bibinfo{howpublished}{\url{https://checkerframework.org/manual/}}.
\newblock
\newblock
\shownote{Accessed: 2019-01-29.}


\bibitem[\protect\citeauthoryear{??}{too}{2019a}]%
        {tool-ep}
 \bibinfo{year}{2019}\natexlab{a}.
\newblock \bibinfo{title}{{E}rror {P}rone}.
\newblock
\newblock
\urldef\tempurl%
\url{http://errorprone.info/}
\showURL{%
\tempurl}
\newblock
\shownote{Accessed: 2019-02-07.}


\bibitem[\protect\citeauthoryear{??}{ep-}{2019a}]%
        {ep-nullderef-check}
 \bibinfo{year}{2019}\natexlab{a}.
\newblock \bibinfo{title}{{Error Prone NullableDereference check}}.
\newblock \bibinfo{howpublished}{\url{https://git.io/fhQkO}}.
\newblock
\newblock
\shownote{Accessed: 2019-01-29.}


\bibitem[\protect\citeauthoryear{??}{ep-}{2019b}]%
        {ep-retmissingnullable-check}
 \bibinfo{year}{2019}\natexlab{b}.
\newblock \bibinfo{title}{{Error Prone ReturnMissingNullable check}}.
\newblock \bibinfo{howpublished}{\url{https://git.io/fhQk3}}.
\newblock
\newblock
\shownote{Accessed: 2019-01-29.}


\bibitem[\protect\citeauthoryear{??}{lib}{2019a}]%
        {lib-guava}
 \bibinfo{year}{2019}\natexlab{a}.
\newblock \bibinfo{title}{Google Core Libraries for {J}ava ({G}uava)}.
\newblock
\newblock
\urldef\tempurl%
\url{https://github.com/google/guava}
\showURL{%
\tempurl}
\newblock
\shownote{Accessed: 2019-02-10.}


\bibitem[\protect\citeauthoryear{??}{too}{2019b}]%
        {tool-eradicate}
 \bibinfo{year}{2019}\natexlab{b}.
\newblock \bibinfo{title}{{I}nfer : {E}radicate}.
\newblock
\newblock
\urldef\tempurl%
\url{https://fbinfer.com/docs/eradicate.html}
\showURL{%
\tempurl}
\newblock
\shownote{Accessed: 2019-01-29.}


\bibitem[\protect\citeauthoryear{??}{jb-}{2019}]%
        {jb-contract}
 \bibinfo{year}{2019}\natexlab{}.
\newblock \bibinfo{title}{{I}ntelliJ {I}DEA @{C}ontract}.
\newblock
\newblock
\urldef\tempurl%
\url{https://www.jetbrains.com/help/idea/contract-annotations.html}
\showURL{%
\tempurl}
\newblock
\shownote{Accessed: 2019-02-07.}


\bibitem[\protect\citeauthoryear{??}{kot}{2019}]%
        {kotlin-lang}
 \bibinfo{year}{2019}\natexlab{}.
\newblock \bibinfo{title}{{Kotlin Programming Language}}.
\newblock \bibinfo{howpublished}{\url{https://kotlinlang.org/}}.
\newblock
\newblock
\shownote{Accessed: 2019-01-29.}


\bibitem[\protect\citeauthoryear{??}{too}{2019c}]%
        {tool-nullaway}
 \bibinfo{year}{2019}\natexlab{c}.
\newblock \bibinfo{title}{{N}ullAway}.
\newblock
\newblock
\urldef\tempurl%
\url{https://github.com/uber/NullAway}
\showURL{%
\tempurl}
\newblock
\shownote{Accessed: 2019-07-01.}


\bibitem[\protect\citeauthoryear{??}{na-}{2019}]%
        {na-eval-repo}
 \bibinfo{year}{2019}\natexlab{}.
\newblock \bibinfo{title}{{Performance Benchmarking of Java Null Safety
  Tools}}.
\newblock
  \bibinfo{howpublished}{\url{https://github.com/subarnob/nullaway-eval}}.
\newblock
\newblock
\shownote{Accessed: 2019-07-01.}


\bibitem[\protect\citeauthoryear{??}{lib}{2019b}]%
        {lib-rxjava}
 \bibinfo{year}{2019}\natexlab{b}.
\newblock \bibinfo{title}{{R}eactive{X}/{R}x{J}ava}.
\newblock
\newblock
\urldef\tempurl%
\url{https://github.com/ReactiveX/RxJava}
\showURL{%
\tempurl}
\newblock
\shownote{Accessed: 2019-02-10.}


\bibitem[\protect\citeauthoryear{??}{sup}{2019}]%
        {supp-data}
 \bibinfo{year}{2019}\natexlab{}.
\newblock \bibinfo{title}{{Supplementary Data}}.
\newblock
  \bibinfo{howpublished}{\url{https://figshare.com/s/a212932795a43c377a3f}}.
\newblock
\newblock
\shownote{Accessed: 2019-02-20.}


\bibitem[\protect\citeauthoryear{??}{swi}{2019}]%
        {swift-lang}
 \bibinfo{year}{2019}\natexlab{}.
\newblock \bibinfo{title}{{Swift Programming Language}}.
\newblock \bibinfo{howpublished}{\url{https://swift.org/}}.
\newblock
\newblock
\shownote{Accessed: 2019-01-29.}


\bibitem[\protect\citeauthoryear{??}{too}{2019d}]%
        {tool-cf}
 \bibinfo{year}{2019}\natexlab{d}.
\newblock \bibinfo{title}{{T}he {C}hecker {F}ramework}.
\newblock
\newblock
\urldef\tempurl%
\url{https://github.com/typetools/checker-framework}
\showURL{%
\tempurl}
\newblock
\shownote{Accessed: 2019-01-29.}


\bibitem[\protect\citeauthoryear{??}{jls}{2019}]%
        {jls9}
 \bibinfo{year}{2019}\natexlab{}.
\newblock \bibinfo{title}{{The Java Language Specification}}.
\newblock
  \bibinfo{howpublished}{\url{https://docs.oracle.com/javase/specs/jls/se9/html/}}.
\newblock
\newblock
\shownote{Accessed: 2019-01-29.}


\bibitem[\protect\citeauthoryear{??}{act}{2019}]%
        {activity-lifecycle}
 \bibinfo{year}{2019}\natexlab{}.
\newblock \bibinfo{title}{{Understand the Activity Lifecycle}}.
\newblock
  \bibinfo{howpublished}{\url{https://developer.android.com/guide/components/activities/activity-lifecycle}}.
\newblock
\newblock
\shownote{Accessed: 2019-01-29.}


\bibitem[\protect\citeauthoryear{??}{gua}{2019}]%
        {guava-collections-null}
 \bibinfo{year}{2019}\natexlab{}.
\newblock \bibinfo{title}{{Using and Avoiding Null Explained}}.
\newblock
  \bibinfo{howpublished}{\url{https://github.com/google/guava/wiki/UsingAndAvoidingNullExplained}}.
\newblock
\newblock
\shownote{Accessed: 2019-01-29.}


\bibitem[\protect\citeauthoryear{Aftandilian, Sauciuc, Priya, and
  Krishnan}{Aftandilian et~al\mbox{.}}{2012}]%
        {DBLP:conf/scam/AftandilianSPK12}
\bibfield{author}{\bibinfo{person}{Edward Aftandilian}, \bibinfo{person}{Raluca
  Sauciuc}, \bibinfo{person}{Siddharth Priya}, {and}
  \bibinfo{person}{Sundaresan Krishnan}.} \bibinfo{year}{2012}\natexlab{}.
\newblock \showarticletitle{Building Useful Program Analysis Tools Using an
  Extensible Java Compiler}. In \bibinfo{booktitle}{\emph{12th {IEEE}
  International Working Conference on Source Code Analysis and Manipulation,
  {SCAM} 2012, Riva del Garda, Italy, September 23-24, 2012}}.
  \bibinfo{pages}{14--23}.
\newblock
\urldef\tempurl%
\url{https://doi.org/10.1109/SCAM.2012.28}
\showDOI{\tempurl}


\bibitem[\protect\citeauthoryear{Blackshear, Gorogiannis, O'Hearn, and
  Sergey}{Blackshear et~al\mbox{.}}{2018}]%
        {DBLP:journals/pacmpl/BlackshearGOS18}
\bibfield{author}{\bibinfo{person}{Sam Blackshear}, \bibinfo{person}{Nikos
  Gorogiannis}, \bibinfo{person}{Peter~W. O'Hearn}, {and} \bibinfo{person}{Ilya
  Sergey}.} \bibinfo{year}{2018}\natexlab{}.
\newblock \showarticletitle{RacerD: compositional static race detection}.
\newblock \bibinfo{journal}{\emph{{PACMPL}}} \bibinfo{volume}{2},
  \bibinfo{number}{{OOPSLA}} (\bibinfo{year}{2018}),
  \bibinfo{pages}{144:1--144:28}.
\newblock
\urldef\tempurl%
\url{https://doi.org/10.1145/3276514}
\showDOI{\tempurl}


\bibitem[\protect\citeauthoryear{Brotherston, Dietl, and
  Lhot{\'{a}}k}{Brotherston et~al\mbox{.}}{2017}]%
        {DBLP:conf/cc/BrotherstonDL17}
\bibfield{author}{\bibinfo{person}{Dan Brotherston}, \bibinfo{person}{Werner
  Dietl}, {and} \bibinfo{person}{Ondrej Lhot{\'{a}}k}.}
  \bibinfo{year}{2017}\natexlab{}.
\newblock \showarticletitle{Granullar: gradual nullable types for Java}. In
  \bibinfo{booktitle}{\emph{Proceedings of the 26th International Conference on
  Compiler Construction, Austin, TX, USA, February 5-6, 2017}}.
  \bibinfo{pages}{87--97}.
\newblock
\urldef\tempurl%
\url{https://doi.org/10.1145/3033019.3033032}
\showDOI{\tempurl}


\bibitem[\protect\citeauthoryear{Calcagno, Distefano, O'Hearn, and
  Yang}{Calcagno et~al\mbox{.}}{2011}]%
        {DBLP:journals/jacm/CalcagnoDOY11}
\bibfield{author}{\bibinfo{person}{Cristiano Calcagno}, \bibinfo{person}{Dino
  Distefano}, \bibinfo{person}{Peter~W. O'Hearn}, {and}
  \bibinfo{person}{Hongseok Yang}.} \bibinfo{year}{2011}\natexlab{}.
\newblock \showarticletitle{Compositional Shape Analysis by Means of
  Bi-Abduction}.
\newblock \bibinfo{journal}{\emph{J. {ACM}}} \bibinfo{volume}{58},
  \bibinfo{number}{6} (\bibinfo{year}{2011}), \bibinfo{pages}{26:1--26:66}.
\newblock
\urldef\tempurl%
\url{https://doi.org/10.1145/2049697.2049700}
\showDOI{\tempurl}


\bibitem[\protect\citeauthoryear{Deutsch}{Deutsch}{1994}]%
        {DBLP:conf/pldi/Deutsch94}
\bibfield{author}{\bibinfo{person}{Alain Deutsch}.}
  \bibinfo{year}{1994}\natexlab{}.
\newblock \showarticletitle{Interprocedural May-Alias Analysis for Pointers:
  Beyond \emph{k}-limiting}. In \bibinfo{booktitle}{\emph{Proceedings of the
  {ACM} SIGPLAN'94 Conference on Programming Language Design and Implementation
  (PLDI), Orlando, Florida, USA, June 20-24, 1994}}. \bibinfo{pages}{230--241}.
\newblock
\urldef\tempurl%
\url{https://doi.org/10.1145/178243.178263}
\showDOI{\tempurl}


\bibitem[\protect\citeauthoryear{Dietl, Dietzel, Ernst, Mu{\c{s}}lu, and
  Schiller}{Dietl et~al\mbox{.}}{2011}]%
        {DietlDEMS2011}
\bibfield{author}{\bibinfo{person}{Werner Dietl}, \bibinfo{person}{Stephanie
  Dietzel}, \bibinfo{person}{Michael~D. Ernst},
  \bibinfo{person}{K{\i}van{\c{c}} Mu{\c{s}}lu}, {and} \bibinfo{person}{Todd
  Schiller}.} \bibinfo{year}{2011}\natexlab{}.
\newblock \showarticletitle{Building and using pluggable type-checkers}. In
  \bibinfo{booktitle}{\emph{ICSE 2011, Proceedings of the 33rd International
  Conference on Software Engineering}}. \bibinfo{address}{Waikiki, Hawaii,
  USA}, \bibinfo{pages}{681--690}.
\newblock
\urldef\tempurl%
\url{https://doi.org/10.1145/1985793.1985889}
\showDOI{\tempurl}


\bibitem[\protect\citeauthoryear{F{\"{a}}hndrich and Leino}{F{\"{a}}hndrich and
  Leino}{2003}]%
        {DBLP:conf/oopsla/FahndrichL03}
\bibfield{author}{\bibinfo{person}{Manuel F{\"{a}}hndrich} {and}
  \bibinfo{person}{K.~Rustan~M. Leino}.} \bibinfo{year}{2003}\natexlab{}.
\newblock \showarticletitle{Declaring and checking non-null types in an
  object-oriented language}. In \bibinfo{booktitle}{\emph{Proceedings of the
  2003 {ACM} {SIGPLAN} Conference on Object-Oriented Programming Systems,
  Languages and Applications, {OOPSLA} 2003, October 26-30, 2003, Anaheim, CA,
  {USA}}}. \bibinfo{pages}{302--312}.
\newblock
\urldef\tempurl%
\url{https://doi.org/10.1145/949305.949332}
\showDOI{\tempurl}


\bibitem[\protect\citeauthoryear{F{\"{a}}hndrich and Xia}{F{\"{a}}hndrich and
  Xia}{2007}]%
        {DBLP:conf/oopsla/FahndrichX07}
\bibfield{author}{\bibinfo{person}{Manuel F{\"{a}}hndrich} {and}
  \bibinfo{person}{Songtao Xia}.} \bibinfo{year}{2007}\natexlab{}.
\newblock \showarticletitle{Establishing object invariants with delayed types}.
  In \bibinfo{booktitle}{\emph{Proceedings of the 22nd Annual {ACM} {SIGPLAN}
  Conference on Object-Oriented Programming, Systems, Languages, and
  Applications, {OOPSLA} 2007, October 21-25, 2007, Montreal, Quebec, Canada}}.
  \bibinfo{pages}{337--350}.
\newblock
\urldef\tempurl%
\url{https://doi.org/10.1145/1297027.1297052}
\showDOI{\tempurl}


\bibitem[\protect\citeauthoryear{Finifter, Mettler, Sastry, and
  Wagner}{Finifter et~al\mbox{.}}{2008}]%
        {DBLP:conf/ccs/FinifterMSW08}
\bibfield{author}{\bibinfo{person}{Matthew Finifter}, \bibinfo{person}{Adrian
  Mettler}, \bibinfo{person}{Naveen Sastry}, {and} \bibinfo{person}{David~A.
  Wagner}.} \bibinfo{year}{2008}\natexlab{}.
\newblock \showarticletitle{Verifiable functional purity in {Java}}. In
  \bibinfo{booktitle}{\emph{Proceedings of the 2008 {ACM} Conference on
  Computer and Communications Security, {CCS} 2008, Alexandria, Virginia, USA,
  October 27-31, 2008}}. \bibinfo{pages}{161--174}.
\newblock
\urldef\tempurl%
\url{https://doi.org/10.1145/1455770.1455793}
\showDOI{\tempurl}


\bibitem[\protect\citeauthoryear{Heule and Garrett}{Heule and Garrett}{2019}]%
        {cf-dataflow-manual}
\bibfield{author}{\bibinfo{person}{Stefan Heule} {and} \bibinfo{person}{Charlie
  Garrett}.} \bibinfo{year}{2019}\natexlab{}.
\newblock \bibinfo{title}{A Dataflow Framework for {J}ava}.
\newblock
\newblock
\urldef\tempurl%
\url{https://checkerframework.org/manual/checker-framework-dataflow-manual.pdf}
\showURL{%
\tempurl}
\newblock
\shownote{Accessed: 2019-02-07.}


\bibitem[\protect\citeauthoryear{Huang, Milanova, Dietl, and Ernst}{Huang
  et~al\mbox{.}}{2012}]%
        {DBLP:conf/oopsla/HuangMDE12}
\bibfield{author}{\bibinfo{person}{Wei Huang}, \bibinfo{person}{Ana Milanova},
  \bibinfo{person}{Werner Dietl}, {and} \bibinfo{person}{Michael~D. Ernst}.}
  \bibinfo{year}{2012}\natexlab{}.
\newblock \showarticletitle{Reim {\&} ReImInfer: checking and inference of
  reference immutability and method purity}. In
  \bibinfo{booktitle}{\emph{Proceedings of the 27th Annual {ACM} {SIGPLAN}
  Conference on Object-Oriented Programming, Systems, Languages, and
  Applications, {OOPSLA} 2012, part of {SPLASH} 2012, Tucson, AZ, USA, October
  21-25, 2012}}. \bibinfo{pages}{879--896}.
\newblock
\urldef\tempurl%
\url{https://doi.org/10.1145/2384616.2384680}
\showDOI{\tempurl}


\bibitem[\protect\citeauthoryear{Loginov, Yahav, Chandra, Fink, Rinetzky, and
  Nanda}{Loginov et~al\mbox{.}}{2008}]%
        {DBLP:conf/issta/LoginovYCFRN08}
\bibfield{author}{\bibinfo{person}{Alexey Loginov}, \bibinfo{person}{Eran
  Yahav}, \bibinfo{person}{Satish Chandra}, \bibinfo{person}{Stephen Fink},
  \bibinfo{person}{Noam Rinetzky}, {and} \bibinfo{person}{Mangala~Gowri
  Nanda}.} \bibinfo{year}{2008}\natexlab{}.
\newblock \showarticletitle{Verifying dereference safety via expanding-scope
  analysis}. In \bibinfo{booktitle}{\emph{Proceedings of the {ACM/SIGSOFT}
  International Symposium on Software Testing and Analysis, {ISSTA} 2008,
  Seattle, WA, USA, July 20-24, 2008}}. \bibinfo{pages}{213--224}.
\newblock
\urldef\tempurl%
\url{https://doi.org/10.1145/1390630.1390657}
\showDOI{\tempurl}


\bibitem[\protect\citeauthoryear{Madhavan and Komondoor}{Madhavan and
  Komondoor}{2011}]%
        {DBLP:conf/oopsla/MadhavanK11}
\bibfield{author}{\bibinfo{person}{Ravichandhran Madhavan} {and}
  \bibinfo{person}{Raghavan Komondoor}.} \bibinfo{year}{2011}\natexlab{}.
\newblock \showarticletitle{Null dereference verification via over-approximated
  weakest pre-conditions analysis}. In \bibinfo{booktitle}{\emph{Proceedings of
  the 26th Annual {ACM} {SIGPLAN} Conference on Object-Oriented Programming,
  Systems, Languages, and Applications, {OOPSLA} 2011, part of {SPLASH} 2011,
  Portland, OR, USA, October 22 - 27, 2011}}. \bibinfo{pages}{1033--1052}.
\newblock
\urldef\tempurl%
\url{https://doi.org/10.1145/2048066.2048144}
\showDOI{\tempurl}


\bibitem[\protect\citeauthoryear{Nanda and Sinha}{Nanda and Sinha}{2009}]%
        {DBLP:conf/icse/NandaS09}
\bibfield{author}{\bibinfo{person}{Mangala~Gowri Nanda} {and}
  \bibinfo{person}{Saurabh Sinha}.} \bibinfo{year}{2009}\natexlab{}.
\newblock \showarticletitle{Accurate Interprocedural Null-Dereference Analysis
  for Java}. In \bibinfo{booktitle}{\emph{31st International Conference on
  Software Engineering, {ICSE} 2009, May 16-24, 2009, Vancouver, Canada,
  Proceedings}}. \bibinfo{pages}{133--143}.
\newblock
\urldef\tempurl%
\url{https://doi.org/10.1109/ICSE.2009.5070515}
\showDOI{\tempurl}


\bibitem[\protect\citeauthoryear{Papi, Ali, Correa~Jr., Perkins, and
  Ernst}{Papi et~al\mbox{.}}{2008}]%
        {PapiACPE2008}
\bibfield{author}{\bibinfo{person}{Matthew~M. Papi}, \bibinfo{person}{Mahmood
  Ali}, \bibinfo{person}{Telmo~Luis Correa~Jr.}, \bibinfo{person}{Jeff~H.
  Perkins}, {and} \bibinfo{person}{Michael~D. Ernst}.}
  \bibinfo{year}{2008}\natexlab{}.
\newblock \showarticletitle{Practical pluggable types for {Java}}. In
  \bibinfo{booktitle}{\emph{ISSTA 2008, Proceedings of the 2008 International
  Symposium on Software Testing and Analysis}}. \bibinfo{address}{Seattle, WA,
  USA}, \bibinfo{pages}{201--212}.
\newblock
\urldef\tempurl%
\url{https://doi.org/10.1145/1390630.1390656}
\showDOI{\tempurl}


\bibitem[\protect\citeauthoryear{Pearce}{Pearce}{2011}]%
        {DBLP:conf/cc/Pearce11}
\bibfield{author}{\bibinfo{person}{David~J. Pearce}.}
  \bibinfo{year}{2011}\natexlab{}.
\newblock \showarticletitle{JPure: {A} Modular Purity System for Java}. In
  \bibinfo{booktitle}{\emph{Compiler Construction - 20th International
  Conference, {CC} 2011, Held as Part of the Joint European Conferences on
  Theory and Practice of Software, {ETAPS} 2011, Saarbr{\"{u}}cken, Germany,
  March 26-April 3, 2011. Proceedings}}. \bibinfo{pages}{104--123}.
\newblock
\urldef\tempurl%
\url{https://doi.org/10.1007/978-3-642-19861-8\_7}
\showDOI{\tempurl}


\bibitem[\protect\citeauthoryear{Pierce}{Pierce}{2002}]%
        {Pierce:2002:TPL:509043}
\bibfield{author}{\bibinfo{person}{Benjamin~C. Pierce}.}
  \bibinfo{year}{2002}\natexlab{}.
\newblock \bibinfo{booktitle}{\emph{Types and Programming Languages}
  (\bibinfo{edition}{1st} ed.)}.
\newblock \bibinfo{publisher}{The MIT Press}.
\newblock
\showISBNx{0262162091, 9780262162098}


\bibitem[\protect\citeauthoryear{Qi and Myers}{Qi and Myers}{2009}]%
        {DBLP:conf/popl/QiM09}
\bibfield{author}{\bibinfo{person}{Xin Qi} {and} \bibinfo{person}{Andrew~C.
  Myers}.} \bibinfo{year}{2009}\natexlab{}.
\newblock \showarticletitle{Masked types for sound object initialization}. In
  \bibinfo{booktitle}{\emph{Proceedings of the 36th {ACM} {SIGPLAN-SIGACT}
  Symposium on Principles of Programming Languages, {POPL} 2009, Savannah, GA,
  USA, January 21-23, 2009}}. \bibinfo{pages}{53--65}.
\newblock
\urldef\tempurl%
\url{https://doi.org/10.1145/1480881.1480890}
\showDOI{\tempurl}


\bibitem[\protect\citeauthoryear{Stein, Clapp, Sridharan, and Chang}{Stein
  et~al\mbox{.}}{2018}]%
        {stein18rx}
\bibfield{author}{\bibinfo{person}{Benno Stein}, \bibinfo{person}{Lazaro
  Clapp}, \bibinfo{person}{Manu Sridharan}, {and} \bibinfo{person}{Bor-Yuh~Evan
  Chang}.} \bibinfo{year}{2018}\natexlab{}.
\newblock \showarticletitle{Safe Stream-Based Programming with Refinement
  Types}. In \bibinfo{booktitle}{\emph{Proceedings of the 33rd {ACM/IEEE}
  International Conference on Automated Software Engineering}}.
\newblock
\urldef\tempurl%
\url{https://doi.org/10.1145/3238147.3238174}
\showDOI{\tempurl}


\bibitem[\protect\citeauthoryear{Summers and M{\"{u}}ller}{Summers and
  M{\"{u}}ller}{2011}]%
        {DBLP:conf/oopsla/SummersM11}
\bibfield{author}{\bibinfo{person}{Alexander~J. Summers} {and}
  \bibinfo{person}{Peter M{\"{u}}ller}.} \bibinfo{year}{2011}\natexlab{}.
\newblock \showarticletitle{Freedom before commitment: a lightweight type
  system for object initialisation}. In \bibinfo{booktitle}{\emph{Proceedings
  of the 26th Annual {ACM} {SIGPLAN} Conference on Object-Oriented Programming,
  Systems, Languages, and Applications, {OOPSLA} 2011, part of {SPLASH} 2011,
  Portland, OR, USA, October 22 - 27, 2011}}. \bibinfo{pages}{1013--1032}.
\newblock
\urldef\tempurl%
\url{https://doi.org/10.1145/2048066.2048142}
\showDOI{\tempurl}


\bibitem[\protect\citeauthoryear{Zibin, Potanin, Li, Ali, and Ernst}{Zibin
  et~al\mbox{.}}{2010}]%
        {DBLP:conf/oopsla/ZibinPLAE10}
\bibfield{author}{\bibinfo{person}{Yoav Zibin}, \bibinfo{person}{Alex Potanin},
  \bibinfo{person}{Paley Li}, \bibinfo{person}{Mahmood Ali}, {and}
  \bibinfo{person}{Michael~D. Ernst}.} \bibinfo{year}{2010}\natexlab{}.
\newblock \showarticletitle{Ownership and immutability in generic Java}. In
  \bibinfo{booktitle}{\emph{Proceedings of the 25th Annual {ACM} {SIGPLAN}
  Conference on Object-Oriented Programming, Systems, Languages, and
  Applications, {OOPSLA} 2010, October 17-21, 2010, Reno/Tahoe, Nevada,
  {USA}}}. \bibinfo{pages}{598--617}.
\newblock
\urldef\tempurl%
\url{https://doi.org/10.1145/1869459.1869509}
\showDOI{\tempurl}


\end{thebibliography}
